\documentstyle[12pt]{article}

\def \a   {\alpha}
\def \ab  {{\alpha \beta}}
\def \ah  {{1\over 2}}
\def \b   {\beta}
\def \be  {\begin{equation}}
\def \bea {\begin{eqnarray}}
\def \bl  {\bigl}
\def \bgl {\biggl}
\def \br  {\bigr}
\def \bgr {\biggr}
\def \c   {\chi}

\def \ct  {\cdot}
\def \cts {\cdots}
\def \d   {\partial}

\def \dt  {\delta}
\def \Dt  {\Delta}
\def \ee  {\end{equation}}
\def \eea {\end{eqnarray}}
\def \ep  {\epsilon}
\def \2e  {2+\epsilon}
\def \et  {\eta}
\def \g   {\gamma}
\def \G   {\Gamma}
\def \h   {\hat}

\def \idxg  {\int d^D x \sqrt g}
\def \idxgh {\int d^D x \sqrt{\hat g}}
\def \inf {\infty}
\def \iR  {\int_M \hat R \ \sqrt{\hat g} \ d^D x}
\def \iK  {\int_{\partial M} \hat K \ \sqrt{\hat \gamma} \ d^{D-1} x}

\def \lts {\ldots}

\def \m   {\mu}
\def \mn  {{\mu \nu}}
\def \n   {\nu}
\def \nab {\nabla}
\def \nn  {\nonumber}
\def \o   {\omega}
\def \ov  {\over}
\def \p   {\phi}

\def \q   {\quad}
\def \qq  {\qquad}
\def \r   {\rho}

\def \ra  {\rightarrow}
\def \rs  {{\rho \sigma}}
\def \s   {\sigma}
\def \sr  {\sqrt}

\def \t   {\tau}
\def \til {\tilde}

\def \vp  {\varphi}

\begin{document}

\setlength{\oddsidemargin}{0cm}
\setlength{\baselineskip}{7mm}

\begin{titlepage}
 \renewcommand{\thefootnote}{\fnsymbol{footnote}}
    \begin{normalsize}
     \begin{flushright}
        TIT-HEP-288 \\
        March 1995
     \end{flushright}
    \end{normalsize}
    \begin{Large}
       \vspace{1cm}
       \begin{center}
         {\LARGE Quantum Gravity with Boundaries
                 near Two Dimensions} \\
       \end{center}
    \end{Large}

  \vspace{10mm}

\begin{center}
           Toshiaki A{\sc ida}\footnote
           {E-mail address : aida@phys.titech.ac.jp}{\sc and}
           Yoshihisa K{\sc itazawa}\footnote
           {E-mail address : kitazawa@phys.titech.ac.jp}\\
      \vspace{1cm}
           {\it Department of Physics,
Tokyo Institute of Technology,} \\
                 {\it Oh-okayama, Meguro-ku, Tokyo 152, Japan}\\
\vspace{15mm}

\end{center}
\begin{abstract}
We evaluate the quantum corrections of the Einstein-Hilbert action
with boundaries in the $2+\ep $ dimensional expansion approach.
We find the Einstein-Hilbert action with boundaries to be
renormalizable to the one loop order.
We compute the geometric entropy beyond the semiclassical
approximation.
It is found that the exact geometric entropy
is related to the string succeptibility
by the analytic continuation in the central charge.
Our results also show that
we can renormalize the divergent quantum
corrections for the Bekenstein-Hawking entropy of blackholes
by the gravitational coupling constant renormalization
beyond two dimensions.
\end{abstract}
\end{titlepage}
\vfil\eject

\section{Introduction}

\hspace*{0.7\parindent}
In quantum gravity, we need to study the influence of boundaries
in many physically interesting questions. We may site the event
horizons in blackhole physics and space-like hyper surfaces
in quantum cosmology. Such questions also arise when we study
the loop amplitudes in two dimensional gravity and open string theory.

In the blackhole spacetime, the existence of the event horizon leads to
the very interesting physics such as the Bekenstein-Hawking entropy
and the Hawking radiation. In the Euclidean blackhole spacetime, the
event horizon is mapped to a point and the spacetime inside the
event horizon simply does not exist. The periodicity of the Euclidean
spacetime (rotation angle around the event horizon) implies that
the system is thermal.

{}From the Minkowski point of view, we need to integrate out the physical
degrees of freedom inside the event horizon. Such an integration leads to
a mixed state. In ref. \cite{BTZ}, it is shown that the blackhole
entropy is given semiclassically
by the Einstein-Hilbert action associated with
the infinitesimal disc around the event horizon.

When we compute the quantum corrections to Bekenstein-Hawking entropy,
it diverges since the divergence of the Einstein-Hilbert action form
arises in the effective action. The difficulty of quantum gravity is the
nonrenormalizability of the theory beyond two dimensions. However
it can be renormalized by the $2+\ep $ dimensional expansion approach.
Furthermore the theory possesses the short distance fixed point with
proper matter contents and consistent quantum gravity theory may be
constructed. Therefore the study of the renormalization of the geometric
entropy in the $2+\ep $ dimensional quantum gravity must be illuminating.

With these physical motivations, we study the renormalization of the
Einstein-Hilbert action with boundaries in the $2+\ep $ dimensions.

\vspace{1.5cm}
\section{$1$-Loop Renormalization}

\hspace*{0.7\parindent}
We shall evaluate the quantum corrections of the Einstein-Hilbert
action with boundaries in the $2+\ep $ dimensional quantum gravity.
Here we adopt the background field method, which gives a gauge
invariant effective action. It is shown in this section
that the divergences are
also of the Einstein-Hilbert action form.
In the first sub-section, we compute the bulk
contributions to the effective action
and
explain our
computational method.
In the following sub-section, we compute the
boundary contributions.

\subsection{Contributions from The Bulk}

\hspace*{0.7\parindent}
We first calculate the quantum corrections of the Einstein-Hilbert
action when a $2+\ep $ dimensional manifold $M$ is not bounded.
They are the bulk contributions proportional to the Einstein action.
As it is expected, we reproduce the well-known result of the conformal
anomaly of two dimensional quantum gravity in the  $2+\ep $
dimensional expansion approach.

\vspace{0.5cm}
Let us consider the action of a free scalar field in a curved space:
\be
- \idxgh \ \ah \ \vp \h \Dt \vp ,
\label{eqn:scalar}
\ee
where $\h \Dt$ is the Laplacian in the curved space. It is defined
in terms of the metric of the curved background $\h g_\mn$ as
\be
\hat \Dt \equiv
{1 \ov \sr{\hat g}} {\d \ov \d x^\m} \sr{\hat g}
\hat g^\mn {\d \ov \d x^\n} .
\label{eqn:lap}
\ee
Here $x^\m $ is a set of local coordinates.

Since we would like to obtain the $1$-loop local divergences,
we only need to consider the short-distance propagation of a particle.
It depends not on the global property of the manifold but on the local
one. So we can adopt the local coordinate method.
When the particle propagates for a very short time, it feels as if
it were moving on the almost flat Euclidean space. Threfore we can
perturb the theory around the flat-space one.

In this paper, we adopt the Riemann's normal coordinates. The
advantage of the method is that we can consider the local property
of the manifold in a manifestly covariant way. In such coordinates,
the Laplacian in the curved background $\h \Dt$ is expanded covariantly
as follows:
\bea
\hat \Dt
& = & ({\d \ov \d u^\m})^2 + {1 \ov 3} \hat R^\m{}_\r{}^\n{}_\s
u^\r u^\s {\d^2 \ov \d u^\m \d u^\n} + {2 \ov 3} \hat R_\m{}^\n
u^\m {\d \ov \d u^\n} + O(\hat R^2)  ,  \nn \\
& \equiv & \Dt + P(u) .
\label{eqn:elap}
\eea
Here $u^\m $ is a set of the geodesic coordinates from a given point
on the manifold $M$ . The Riemann and Ricci curvatures are evaluated at
the origin of the normal coordinates $u^\m = 0$. $\Dt$ and $P(u)$ denote
the flat space Laplacian and the perturbation around it, respectively.

In general, we can obtain the $1$-loop effective action by integrating
the quadratic terms of the action.
\bea
\G_{\rm matter}
& = & \ah \ log Det \ \bl( {\Dt + P \ov \Dt} \br) \ , \nn \\
& = & \ah \ Tr [ log \{- (\Dt+P) \} - log(- \Dt) ] \ ,
\label{eqn:ea}
\eea

We can reexpress the above by introducing a proper time $\t $ as
follows \cite{A}:
\be
\G_{\rm matter} =
- \ah \int_{-\inf}^\inf d^D x \sr{\h g} \int_0^\inf {d \t \ov \t}
[ <x| e^{-\t \{ -(\Dt+P) \} } |x> - <x| e^{-\t (-\Dt)} |x> ] .
\ee
Here, $\h G(x_1,x_2;\t) \equiv <x_1 | e^{-\t \{ -(\Dt+P) \} } |x_2 >$ is
called a heat kernel. This is because it is the Green's function of the
heat
equation
\be
\bl\{ {\d \ov \d \t} - (\Dt_{x_1} + P(x_1)) \br\} \h G(x_1,x_2;\t)
= \dt(\t ) \dt^{(D)} (x_1-x_2) /\sr{\h g(x_2)} .
\ee
On the other hand, $G(x_1,x_2;\t) \equiv <x_1| e^{-\t (-\Dt ) } |x_2>$ is
the flat space Green's function, satisfying
\be
\bl( {\d \ov \d \t} - \Dt_{x_1} \br) G(x_1,x_2;\t)
= \dt(\t ) \dt^{(D)} (x_1-x_2)/\sr{\h g (x_2)} .
\ee
Its solution is easily obtained by Fourier transformations.
\bea
G(x_1,x_2;\t)
& = & {1 \ov \sr{\h g (x_2)} } \int_{-\inf}^\inf {d^D p \ov (2 \pi)^D}
e^{-\t p^2} e^{i p \ct (x_1-x_2)}  \nn \\
& = & {1 \ov \sr{\h g (x_2)} }
{1 \ov (4 \pi \t)^{D \ov 2} } e^{- {(x_1-x_2)^2 \ov 4 \t} }
\eea
We can obtain $\h G$ perturbatively, in terms of the flat space
Green's function $G$.
\be
\h G = G + G P G + \lts .
\ee
As a result, we only have to evaluate the following integration.
\bea
\G_{\rm matter} & = &
- \ah \int_{-\inf}^\inf d^D x \sr{\h g} \int_0^\inf {d \t \ov \t} \nn \\
& & \times \bgl[ \int_0^\inf d \t_1 d \t_2 \dt( \t -\t_1 -\t_2 )
\int_{-\inf}^\inf d^D x' \ G(x,x';\t_1 ) P(x') G(x',x;\t_2 ) \ +
\lts \bgr] .
\eea
The calculation of the above is straightforward, where it is convenient
to choose $x$ as the origin of the normal coordinate expansion. We find
the result of the $1$-loop divergence of a real scalar field
in $ D = 2 + \ep$ dimensions as
\be
\G_{\rm matter} \simeq -{1 \ov 24 \pi \ep} \iR .
\ee

\vspace{1cm}
Next we shall consider the gravitational and ghost fields.

We adopt the parametrization of the gravitational degrees of freedom
by Kawai, Kitazawa and Ninomiya \cite{2e}, which singles out the conformal
mode of the metric $\p$:
\bea
g_\mn & \equiv & \til g_\mn e^{-\p} ,  \nonumber \\
      & \equiv & \hat g_{\m \r} (e^h)^\r{}_\n e^{-\p} ,
\eea
where $\hat g_\mn$ is the background metric and $h_\mn$
is a traceless symmetric matrix $(\hat g^\mn h_\mn = 0)$.
The tensor indices are raised or lowered by the background metric.
In such a parametrization, the Einstein action near two
dimensions becomes
\bea
{\m^\ep \ov G} \idxg R
& = & {\m^\ep \ov G} \idxgh \h R  \nn \\
& + & {\m^\ep \ov G} \idxgh \bl\{
{1 \ov 4} \h \nab_\r h^\m{}_\n \h \nab^\r h^\n{}_\m
+ \ah \h R^\s{}_{\m \n \r } h^\r{}_\s h^{\m \n }  \nn \\
& - & {\ep \ov 4} (D-1) \h g^{\m \n } \d_\m \p \d_\n \p
+ {\ep^2 \ov 8} \p^2 \h R + {\ep \ov 2} \p h^\m{}_\n \h R^\n{}_\m   \\
& - & \ah \h \nab_\m h^\m{}_\r \h \nab_\n h^{\n \r }
+ {\ep \ov 2} \p \h \nab_\m \h \nab_\n h^\mn
\br\} + \lts , \nn
\eea
where $G, \ \m $ are the gravitational coupling constant and
a renormalization scale to define it, respectively.

In order to cancel the last two terms, we adopt a Feynman-like gauge:
\be
{\m^\ep \ov G} \idxgh \ah
\bl( \h \nab_\m h^\m{}_\r + {\ep \ov 2} \d_\r \p \br)
\bl( \h \nab_\n h^{\n \r } + {\ep \ov 2} \d^\r \p \br) .
\ee

The change of the metric under the general coordinate transformation is
\be
\dt g_\mn = \d_\m \ep^\r g_{\r \n } + g_{\m \r } \d_\n \ep^\r
+ \ep^\r \d_\r g_\mn \ .
\ee
It leads the gauge transformations of $h^\m{}_\n $ and $\p $ fields as:
\bea
\dt h^\m{}_\n & = &
\h \nab^\m \ep_\n + \h \nab_\n \ep^\m
- {2 \ov D} \h \nab_\r \ep^\r \dt^\m{}_\n + \lts ,  \nn \\
\dt \p & = & \ep^\m \d_\m \p
- {2 \ov D} \h \nab_\m \ep^\m + \lts \ \ .
\eea
Following the standard procedure, we find the ghost action to be
\be
{\m^\ep \ov G} \idxgh
( \bar \et_\m \h \nab_\n \h \nab^\n \et^\m
- \h R^\m{}_\n \bar \et_\m \et^\n
- {\ep \ov 2} \d_\n \p \h \nab^\m \bar \et_\m \et^\n
+ \lts ) .
\ee

In this way, we find the quadratic terms needed for the $1$-loop
calculations in the background gauge.
\bea
& & {\m^\ep \ov G} \idxgh \bl[
{1 \ov 4} (\dt^\m{}_\r \dt^\n{}_\s - {1 \ov D} \hat g^\mn \hat g_\rs)
\hat \nab_\a h_\mn \hat \nab^\a h^\rs
+ \ah \hat R^\s{}_{\mn\r} h^\r{}_\s h^\mn  \nonumber \\
& & \qquad \qquad \quad \ -{\ep \ov 8} D \hat g^\mn \d_\m \p \d_\n \p
+{\ep^2 \ov 8} \hat R \p^2      \\
& & \qquad \qquad \quad \ -\dt^\m{}_\n \hat \nab_\a \bar \et_\m
\hat \nab^\a \et^\n
-\hat R^\m{}_\n \bar \et_\m \et^\n \br] .  \nonumber
\eea
As a result, we can evaluate the Green's functions for $h_\mn$ , $\p$
and ghost fields.
\bea
& & \bl[ \h I^\mn{}_{, \ab } (x_1) {\d \ov \d \t}
- \{ \h I^\mn{}_{, \ab } (x_1) \h \Dt_{x_1}
+ 2 I^\mn{}_{, \gamma\delta } (x_1)
\h R^\g{}_\a{}^\dt{}_\b (x_1) \} \br]
\h G^\ab{}_{, \rs } (x_1,x_2;\t ) \nn \\
& & \qq \qq \qq \qq \qq \qq \qq \qq \
= \ \h I^\mn{}_{, \rs } (x_2) \ \dt (\t ) \dt^{(D)} (x_1-x_2)
/ \sr{\h g (x_2)} \ , \nn \\
& & \bl[ {\d \ov \d \t} -( \h \Dt_{x_1} + {\ep \ov D} \h R(x_1) ) \br]
\h G_\p (x_1,x_2;\t )
\ = \ \dt (\t ) \dt^{(D)} (x_1-x_2) / \sr{\h g (x_2)} \ ,
\label{eqn:deq} \\
& & \bl[ \dt^\m{}_\r {\d \ov \d \t}
-( \dt^\m{}_\r \h \Dt_{x_1} - \h R^\m{}_\r (x_1) ) \br]
\h G^\r{}_\n (x_1,x_2;\t )
\ = \ \dt^\m{}_\n \ \dt (\t ) \dt^{(D)} (x_1-x_2)
/ \sr{\h g (x_2)} , \nn
\eea
where
$\h I^\mn{}_{,\ab} (x) = \ah \dt^\m{}_\a \dt^\n{}_\b + \ah \dt^\m{}_\b
\dt^\n{}_\a -{1\ov D} \h g^\mn (x) \h g_\ab (x)$
is the identity for the traceless symmetric tensors in a $D$ dimensional
curved space. We have normalized the heat kernels so that the coefficients
of the Laplacians are equal to one. It is allowed to do so since we
consider the normalization independent ratio as in the eqn. (\ref{eqn:ea}).

We note that the boundary terms do not appear in the
eqs. (\ref{eqn:deq}), since we assume that $h_\mn$ , $\p$ and ghost
fields fall off rapidly enough at $x \ra \inf$.
In the next sub-section, we consider the case where the manifold $M$ has
a boundary. It will be seen that we also obtain the
eqs. (\ref{eqn:deq}). This is because the boundary term
which arises when we take the
variation of the Einstein term cancells out that of the extrinsic
curvature term.

In a similar way to a scalar field case, we can obtain
the heat kernel for $h_\mn$ , $\p$ and ghost fields
in a curved background as
$\hat G^\mn{}_{,\rs}$ , $\h G_\p$ and $\hat G^\m{}_\n$
respectively. Here, it is convenient to choose $x_2$ as the origin of the
normal coordinates $u^\m = 0$ and to assign the normal coordinates $u^\m$
to $x_1$. We note that the geometrical quantities evaluated at $u$ are
expressed in terms of those evaluated at the origin $u^\m = 0$, as follows:
\bea
\h g_\mn (u) & = &
\dt_\mn - {1 \ov 3} \h R_{\m \r \n \s} (0) u^\r u^\s
+ \lts \ , \nn \\
\h R_{\mn \rs } (u) & = &
\h R_{\mn \rs } (0)
+ \lts \ , \nn \\
\h R_\mn (u) & = &
\h R_\mn (0) + \lts \ , \\
\h R (u) & = &
\h R (0) + \lts \ . \nn
\eea
Here, the dots express the higher order terms, which are unnecessary for
us to calculate divergent contributions for the effective action.
It is important to note that the Riemann and Ricci curvatures at $u$ are
equal to those at the origin up to this order. In the following,
we simply express $\h R_{\mn \rs} (0) , \cts$ as $\h R_{\mn \rs} , \cts$.
In terms of them, we get the following $1$-loop divergences
from $\p$ , $h_\mn$ and ghost fields.
\begin{eqnarray}
\G_\p & \simeq & -{1 \ov 24 \pi \ep} \iR , \nn \\
\G_h & \simeq & \bl( -{2 \ov 24 \pi \ep} + {1 \ov 2 \pi \ep} \br) \iR , \\
\G_{\rm ghost} & \simeq &
\bl( -{-4 \ov 24 \pi \ep} + {1 \ov 2 \pi \ep} \br) \iR . \nn
\end{eqnarray}
The conformal mode gives the identical contribution with that
of a scalar field. It is due to the fact that, for the conformal mode,
the perturbation proportional to $\h R$ is $O(\ep )$ smaller than the
kinetic term.

Consequently, we obtain the total $1$-loop divergences of the theory
from the bulk:
\begin{equation}
\G_{\rm div.} = {25-c \ov 24 \pi \ep} \iR .
\label{eqn:bdiv}
\end{equation}
We need to add the counter term to cancel this
divergence. However the counter term breaks the conformal invariance
of the otherwise conformally invariant theory.
This is the origin of the well known conformal anomaly of two
dimensional quantum gravity in our approach.

\vspace{2cm}
\subsection{Contributions from The Boundary}

\hspace*{0.7\parindent}
In this sub-section,
we consider a $D$-dimensional manifold $M$ bounded by
a $(D-1)$-dimensional smooth boundary $\d M$ .
The corrections to the $1$-loop divergence (\ref{eqn:bdiv}) due to
the existence of the boundary is proportional to the extrinsic
curvature of the manifold. The combination of the bulk and boundary
contributions turns out to be of the Einstein-Hilbert action form.

\vspace{0.5cm}
In the vicinity of the boundary, it is convenient to specialize the
coordinates of an interior point $P$
by a new coordinate set $(w,x^i) \ [i=1,\lts ,D-1]$.
The first coordinate $w$ is the geodesic distance
from $P$ to $\o$, which is the projection of $P$ on the boundary,
and the other $D-1$ coordinates $x^i$ characterize the position
of $\o$ on the boundary. We further specialize the coordinates $x^i$
of $ \ \o$, using a set $y^i$ of Riemann's normal coordinates from a given
point $\o_0$ on the boundary.

In this set of coordinates, the metric has only the diagonal components,
and the Laplacian (\ref{eqn:lap}) is expanded as:
\be
\h \Dt = {\d^2 \ov \d w^2} + \bl( {\d \ov \d y^i} \br)^2
- {D-1 \ov R} {\d \ov \d w} + 2w \sum_{i=1}^{D-1} {1 \ov R_i}
\bl( {\d \ov \d y^i} \br)^2 + \lts ,
\label{eqn:blap}
\ee
where $R_i$ are the main curvature radii of the boundary at $\o_0$,
and $R$ is the mean curvature defined by
\be
{1 \ov R} \equiv {1 \ov D-1} \sum_{i=1}^{D-1} {1 \ov R_i} .
\ee

As it will be  seen in the following, the corrections of the unperturbed
Green's function $G(x_1,x_2;\t )$ due to the existence of a boundary are
exponentially damped when $x_1$ and $x_2$ move away from the boundary.
So we only have to consider the vicinity of the boundary to evaluate
the influence of it. Since we would like to obtain the local divergences,
we need not consider the long-distance propagation of a particle. If the
particle propagates near the boundary for a very short time, it believes
as if the boundary were flat. Therefore we can well
approximate the unperturbed Green's function by that of a half Euclidean
space:
\bea
G(x_1,x_2;\t )
& = & G_0 (w_1,y_1;w_2,y_2;\t ) \mp G_0 (w_1,y_1;-w_2,y_2;\t ) , \nn \\
& \equiv & G_0 (x_1,x_2;\t ) + G_1 (x_1,x_2;\t ) ,
\eea
where $G_0$ is the free space Green's function, and the signs $-$ and $+$
correspond to the Dirichlet and Neumann boundary conditions respectively.
This is because the signs $-$ and $+$ make $G(x_1 ,x_2;\t )$ to be
anti-symmetric and symmetric respectively under the inversion of the signs
of the coordinate $w$. $G_1 $ is the correction of the unperturbed
Green's function due to the existence of the boundary.
The explicit forms of $G_0$ and $G_1$ are given by
\bea
G_0 (w_1,y_1;w_2,y_2;\t ) & = &
{1 \ov \sr{\h g (x_2)} } \int {dq d^{D-1} p \ov (2\pi)^D } e^{-\t (q^2 + p^2 )}
e^{iq(w_1 - w_2)} e^{ip(y_1 - y_2)} , \nn \\
& = & {1 \ov \sr{\h g (x_2)} }
{1 \ov (4 \pi \t )^{D \ov 2} } e^{- {(w_1 - w_2)^2 \ov 4 \t} }
e^{- {(y_1 - y_2)^2 \ov 4 \t} } , \nn \\
G_1 (w_1,y_1;w_2,y_2;\t ) & = &
\mp {1 \ov \sr{\h g (x_2)} } \int {dq d^{D-1} p \ov (2\pi)^D } e^{-\t (q^2 +
p^2 )}
e^{iq(w_1 +  w_2)} e^{ip(y_1 - y_2)} , \\
& = & \mp {1 \ov \sr{\h g (x_2)} }
{1 \ov (4 \pi \t )^{D \ov 2} } e^{- {(w_1 + w_2)^2 \ov 4 \t} }
e^{- {(y_1 - y_2)^2 \ov 4 \t} } . \nn
\eea
We note that the free space Green's function $G_0 $ has the
translational invariance in the flat space limit,
while the correction $G_1 $ does not in the
direction perpendicular to the boundary. The $G_1$ decreases
exponentially as the distance from the boundary increases. Since only
the $w=0$ cotributes in the large momentum limit($ \ q,p \ra \inf$) or
the short time limit($ \ \t \ra 0$), we obtain the divergences due
to the presence of the
boundary from $G_1 $.

In terms of this unperturbed Green's function, we can extract
the corrections of Green's function due to the existence of the boundary
by subtracting the perturbative expansions of the free space
Green's function from those of the Green's function of the bounded space
\cite{BB}.
\bea
\dt \h G(x_1,x_2;\t ) & = & G_1 (x_1,x_2;\t )
+ \int_0^\inf d \t_1 d \t_2 \dt( \t -\t_1 -\t_2 )
\int_0^\inf dw' \int_{-\inf}^\inf d^{D-1} y' \sr{\h g(x')}  \nn \\
& & \times
\bl[ - G_0 (x_1,\bar x';\t_1) \ P(\bar x') \ G_0 (\bar x' ,x_2;\t_2)
+ G_0 (x_1,x';\t_1) \ P(x') \ G_1 (x',x_2;\t_2) \label{eqn:cor} \\
& & \q + G_1 (x_1,x';\t_1) \ P(x') \ G_0 (x',x_2;\t_2)
+ G_1 (x_1,x';\t_1) \ P(x') \ G_1 (x',x_2;\t_2) \br] + \lts , \nn
\eea
where $x = (w,y^i), \ \bar x = (-w,y^i)$ and $P(x')$ is
the perturbation given by the $3$rd and $4$th terms
of the r.h.s. of (\ref{eqn:blap}) .

We can now evaluate the $1$-loop divergences from the boundary
due to a free scalar field, using $\dt \h G$ ,
\bea
\dt \G_{\rm matter} & = &
- \ah \int_0^\inf dw \int_{-\inf}^\inf d^{D-1} y \sr{\h \g}
\int_0^\inf {d \t \ov \t}
[ \dt \h G(w,y;w,y;\t ) -  G_{1}(w,y;w,y;\t ) ] , \nn \\
& \simeq &  {1 \ov 12\pi\ep} \iK .
\eea
$\h K$ and $\h \g$ are the extrinsic curvature of the boundary and
the restriction of the metric to the boundary, respectively. $\h K$ is
defined in terms of the inward unit normal vector $n^i$
as $\h K = \h \g_j{}^i \h \nab_i n^j$. We have used the relation between
the mean curvature and the extrinsic curvature:$ \ (D-1)/R =  \h K$.
In the both Dirichlet and Neumann boundary conditions, we obtain
the above result. Therefore the sum of the bulk and boundary
contributions due to a free scalar field results in
\be
\G_{\rm matter} \simeq
-{1 \ov 24\pi\ep} \bl( \iR - 2 \iK \br) .
\ee
This combination of the scalar curvature $\h R$ and the
extrinsic curvature $\h K$ reminds us of the Gauss-Bonnet theorem:
\be
\c (M) \ = \ -{1 \ov 4\pi}
\bl( \int_M \h R \ \sr{\h g} \ d^2 x
- 2 \int_{\d M} \h K \ \sr{\h \g } \ dx \br) ,
\ee
which gives a topological invariant of two dimensional
manifolds: the Euler number. Indeed, the classical action
for the matter is conformally invariant in two dimensions. This is
the reason why we have obtained the divergence which becomes the
topological invariant in the two dimensional limit.

\vspace{1cm}
Next we shall evaluate the corrections from gravitational and
ghost fields. When a manifold is bounded by a $(D-1)$-dimensional
sub-manifold, we have to add a surface term to the action
with the Dirichlet boundary condition to obtain the Einstein's field
equation as the classical equation of the action \cite{W}.
\be
I \ = \ {\m^\ep \ov G}
\bl[ \int_M  R \ \sr{\h g} \ d^D x
- 2 \int_{\d M} K \ \sr{\h \g} \ d^{D-1} x
+ (\rm{gauge \ fixing \ and \ ghost \ terms}) \br] .
\label{eqn:ba}
\ee
Here the linear terms of $h_\mn$ fields and the conformal mode are
dropped since the background fields satisfy Einstein's field equation,
which is obtained by considering the variation of the action with
the Dirichlet boundary conditions for $h_\mn$ fields and the conformal
mode. We note that the heat equations for $h_\mn $ fields and the conformal
mode are the same as those in the unbounded manifold (\ref{eqn:deq})
respectively, due to the surface term and the Dirichlet boundary
condition.

Using the Green's function (\ref{eqn:cor}), we can calculate the boundary
contribution from $h_\mn$ fields:
\be
\dt \G_h \simeq \bl( {2 \ov 24\pi\ep} - {1 \ov 2\pi\ep} \br) \ct
2 \iK .
\ee
The second part of the above expression comes from the
tad pole divergence at the boundary.
We have used the fact that the
Gauss-Bonnet combination is free from the boundary
contribution.
The divergences from the conformal
mode is identical to that from a free scalar field as in the bulk
contribution.

For ghost fields, we should also choose the Dirichlet boundary condition.
To see this, it is convenient to adopt the normal coordinates
explained in the above. In those coordinates, we can easily see that
$\d_\m h^\mn = 0$ on $\d M$ since $h_\mn$ fields are diagonalized
as $h_{00}=0 \ , \ h_{ij} = - {2w \ov R_i} \dt_{ij} + \lts (i,j = 1, \lts
,D-1)$ .
So we obtain $\bar \et_\m = 0$ on $\d M$ from the following relation
between the gauge fixing and ghost terms.
\be
\dt_B (\bar \et_\n \d_\m h^\mn ) \ = \
\ah (\d_\m h^\mn )^2 - \bar \et_\n \d_\m (\dt_B h^\mn ) ,
\ee
where $\dt_B$ denotes the BRS transformation.
Choosing the Dirichlet boundary condition, we also obtain the same
heat equation for ghost fields as (\ref{eqn:deq}).

In a similar fashion to the $h_\mn$ field's case, we can calculate
the boundary contribution from the ghost field.
\be
\dt \G_{\rm ghost} =
\bl( {-4 \ov 24\pi\ep} - {1 \ov 2\pi\ep} \br) \ct 2 \iK .
\ee

We note that the sums of the bulk and boundary divergences
of $\p$, $h_\mn$ and ghost fields take the Euler class forms
in the two dimensional limit, respectively.
\bea
\G_\p & \simeq & -{1 \ov 24\pi\ep} \bl( \iR - 2 \iK \br) , \nn \\
\G_h  & \simeq & \bl( -{2 \ov 24 \pi \ep} + {1 \ov 2 \pi \ep} \br)
\bl( \iR - 2 \iK \br), \\
\G_{\rm ghost} & \simeq &
\bl( -{-4 \ov 24 \pi \ep} + {1 \ov 2 \pi \ep} \br)
\bl( \iR - 2 \iK \br). \nn
\eea

Consequently we obtain the total $1$-loop divergence from $c$ copies
of sclar fields, $h_\mn$, $\p $ and ghost fields:
\be
\G_{\rm div.} \ = \
{25-c \ov 24\pi\ep} \bl( \iR - 2 \iK \br) .
\ee
These divergent terms are the extension of the result in the unbounded
case (\ref{eqn:bdiv}). They have the form proportional to the
Einstein-Hilbert action with boundaries.
We note that they are also propotional to the Euler class
in the two dimensional limit. It is naturally expected
since only the BRS trivial parts of the action (\ref{eqn:ba}) break
the conformal invariance in two dimensions.

The bare action with the counter term is
\be
I_0 \ = \ {1\ov G_0}
\bl[ \int_M  R \ \sr{ g} \ d^D x
- 2 \int_{\d M} K \ \sr{ \g} \ d^{D-1} x
 \br] .
\label{eqn:bare}
\ee
where the bare gravitational coupling is
${1\over {G_0}} = {{\mu}^\epsilon} ({1\over G} - {25-c\over {24\pi\epsilon}})$.
Therefore we can
compensate the divergence by renormalizing the gravitational coupling
constant and need not introduce an additional parameter to the theory.

\vspace{1.5cm}
\section{ Conclusions }

\hspace*{0.7\parindent}
We have evaluated the quantum corrections of the Einstein-Hilbert action
with boundaries in the $2+\ep $ dimensional expansion approach.
The $2+\ep $
dimensional manifold $M$ is assumed to have
a $1+\ep $ dimensional smooth boundary.

We have imposed the Dirichlet or Neumann boundary conditions for the
matter fields. When we consider the quantum fluctuations around the
classical background $\h g_\mn $, we are led to choose the Dirichlet
boundary conditions for $h_\mn $ field and the conformal mode. This is
because the equation of motion for the classical background becomes
the Einstein's field equation only when we choose the Dirichlet boundary
conditions for $h_\mn $ field and the conformal mode. It is found from
the BRS formalism that we should also impose the Dirichlet boundary
condition for ghost fields.

We have studied the $1$-loop corrections of the Einstein-Hilbert action
with boundaries from the matter, $h_\mn$, $\p $ and ghost fields.
The divergences are also of the Einstein-Hilbert action form
with boundaries. Therefore the divergences are removed by the
renormalization of the gravitational coupling constant.

Our result has an application to the Bekenstein-Hawking entropy of
blackholes. As mentioned in the introduction, the entropy of blackholes
is given by the Einstein-Hilbert action associated with the
infinitesimal discs.
The $2+\ep $ dimensional expansion approach shows
that one loop divergence of the Einstein-Hilbert action form arises.
Therefore
the quantum corrections for the blackhole entropy
are also divergent.
However it is also clear that we can obtain the
finite quantum corrections for the blackhole entropy by renormalizing
the gravitational coupling constant.

The Euclidean blackhole spacetime in $D$ dimensions
has the topology $R^2 \times S^{\epsilon}$
where $S^{\epsilon}$ is a $\epsilon$ dimensional
sphere.
The blackhole entropy is the Euler class of a small disk
centered at the horizon multiplied by the area $A_{\epsilon}$
of the $S^{\epsilon}$ there\cite{BTZ}:
\be
S_{BH} = {4\pi \over G_0} A_{\epsilon},
\ee
which becomes the standard formula if we adopt
the standard convention
$G_0 \rightarrow 16\pi G_0$.
The renormalization group improved semiclassical entropy
of the blackhole is
\be
S_{BH} = {4\pi{\mu}^{\epsilon}\over G(\mu)} A_{\epsilon},
\ee
where we have replaced the Newton constant by the running
coupling constant. It is natural to choose the renormalization
scale $\mu$ to match the blackhole scale
such that $\mu^{\epsilon} A_{\epsilon} = 1$.
Then the renormalized blackhole entropy changes with
the scale of the Blackhole as
\be
\mu{d\over {d\mu}} S_{BH} = -(\epsilon  -{25-c\over {24\pi}}G)
S_{BH},
\label{eqn:renent}
\ee
where we have used the renormalization group equation for $1/G$
\cite{2e}.

In the literature, the entropy of the blackholes
and closely related geometric entropy have
been studied\cite{CW,SU,HLW}.
Our results are certainly consistent with these results.
In particular the conformal mode dependence of the
geometric entropy is studied in \cite{HLW}.
In our approach, the conformal mode
dependence of the geometric entropy comes
from the counter term. It is the only source
of the conformal mode dependence in two dimensions
since the tree action is conformally invariant
in two dimensional limit.

Let us consider the geometric entropy of a manifold
with a closed boundary.
The variation of the entropy with respect to the
scale transformation is:
\begin{eqnarray}
\delta S &=&
-\delta I_0  \nonumber \\
&=&
\delta \phi ({\epsilon \over {2G}} -{25-c\over{48\pi}})
\mu ^{\epsilon}
\bl[ \int_M  R \ \sr{\h g} \ d^D x
- 2 \int_{\d M}  K \ \sr{\h \g} \ d^{D-1} x
 \br]
\end{eqnarray}
This formula is consistent with the renormalization group
equation (\ref{eqn:renent}).
By taking the two dimensional limit, we obtain
\be
{\delta S \over {\delta \phi}}
= {25-c\over 12} .
\ee

The constant mode of $\phi$ is related to the area of
a disc as
\be
\int_M  exp(-\alpha \phi) d^2 x = A,
\ee
where we also have to renormalize the cosmological
constant operator in fully quantum theory\cite{DDK}.
Hence we find $\phi \sim -{1\over \alpha} logA$.
The requirement of the conformal invariance determines
$\alpha = {25-c\over 12} -{\sqrt{(1-c)(25-c)}\over 12}$ for
$c<1$.
For $c>25$, $\alpha = {25-c\over 12} +{\sqrt{(c-1)(c-25)}\over 12}$.
These formulas possess the correct semiclassical limit for
large $|c|$. They are related by the analytic continuation in $c$.
We find the scale dependence of the exact
geometric entropy as
\be
{\delta S \over {\delta logA}}
= -\alpha {25-c\over 12} .
\label{eqn:entropy}
\ee
This result agrees with \cite{HLW} in the leading
order of $c$.
Here again we have a difficulty to interpret the theory
for $1<c<25$.

Comparing to the semiclassical results,
the physical meaning of (\ref{eqn:entropy}) is much more
transparent for $c<1$.
In two dimensions, the Gauss-Bonnet action
is topological. Therefore the induced Liouville action
represents the entropy of the theory.
Our results has followed from the
same quantum effect.
It can be interpreted as the quantum entropy
in association with the two dimensional disc
with a fixed area.
In fact it is nothing but the string succeptibility
for $c<1$.
The geometric entropy for $c>25$ can be obtained
by the analytic continuation in $c$.
The difficulty to quantize the theory with
$c>25$ in Euclidean spacetime is the conformal mode instability.
On the other hand the entropy is defined in
Euclidean spacetime.
The conformal mode
instability always exists beyond two dimensions
in the semiclassical regime.

Therefore it is likely that the concept of the blackhole
entropy and Hawking radiation are valid only in the
semiclassical approximation.
Although we have studied geometric entropy beyond
the semiclassical approximation, we have to contemplate
the physical implications of our investigations.
Nevertheless we expect that the whole physical
picture holds as
a very good approximation
in the weak coupling regime.
Then it certainly makes sense to
ask what is the temperature of such a quasithermal object
as a blackhole.
We expect that our results are valid in such a physical
interpretation.


\end{document}